\documentstyle[epsf]{l-aa}
\def\simgr{\,\hbox{\hbox{$ > $}\kern -0.8em \lower 1.0ex\hbox{$\sim$}}\,}
\def\simle{\,\hbox{\hbox{$ < $}\kern -0.8em \lower 1.0ex\hbox{$\sim$}}\,}
\newcommand{\Lsun}{{\mathrm{L}_{\odot}}}
\newcommand{\Msun}{{\mathrm{M}_{\odot}}}
\newcommand{\Rsun}{{\mathrm{R}_{\odot}}}
\newcommand{\km}{{\mathrm{km}}}
\newcommand{\Sec}{{\mathrm{s}}}
\newcommand{\yr}{{\mathrm{yr}}}
\newcommand{\K}{{\mathrm{K}}}

\newcommand{\kms}{{\km\,\Sec^{-1}}}
\newcommand{\Mbol}{{\mathrm{M}_{\mathrm{bol}}}}
\newcommand{\Mag}{{\mathrm{mag}}}
\newcommand{\mso}{{\Msun}}
\newcommand{\msol}{{\Msun}}

\newcommand{\rso}{{\Rsun}}
\newcommand{\lsol}{{\Lsun}}

\def\msoy{\,\mso\,\yr^{-1}}

\newcommand{\Ys}{{Y_{\mathrm{s}}}}
\newcommand{\aMLT}{{\alpha_{\mathrm{MLT}}}}
\newcommand{\aS}{{\alpha_{\mathrm{sem}}}}
\newcommand{\Teff}{{T_{\mathrm{eff}}}}
\newcommand{\MZAMS}{{M_{\mathrm{ZAMS}}}}
\newcommand{\Hef}{{\Ys}}
\newcommand{\MHe}{{M_{\mathrm{He}}}}
\newcommand{\MCO}{{M_{\mathrm{C/O}}}}
\newcommand{\tKH}{{\tau_{\mathrm{KH}}}}

\newcommand{\thyd}{{\tau_{\mathrm{H}}}}
\newcommand{\thel}{{\tau_{\mathrm{He}}}}
\newcommand{\tred}{{\tau_{\mathrm{red}}}}

\newcommand{\Mz}{{\MZAMS}}
\newcommand{\te}{{\Teff}}

\begin{document}
\thesaurus{08.19.3, 08.15.1, 08.16.1, 08.22.3, 08.19.4}
%\title{Pulsations in cool supernova progenitors with high L/M ratio}
 \title{Pulsations in red supergiants with high L/M ratio}
 \subtitle{Implications for the stellar and circumstellar structure of 
    supernova progenitors}

\author{
A.Heger \inst {1,2} 
\and L. Jeannin \inst {3} 
\and N. Langer \inst {1,4}
\and I. Baraffe \inst {3}
}
\institute{
Max-Planck-Institut f\"ur Astrophysik, 
D-85740 Garching, Germany
\and
UCO/LICK Observatory, University of California, Santa Cruz, CA 95064, 
U.S.A.
\and
 Centre de Recherche Astronomique de Lyon  
(UMR 5574, CNRS), Ecole Normale Sup\'erieure de Lyon, 46 all\'ee d'Italie, 
F-69364 Lyon Cedex 07, France
\and
Institut f\"ur Theoretische Physik und Astrophysik, Universit\"at Potsdam,
D-14415 Potsdam, Germany
%
%$3$ Astronomy Board of studies, UC Santa Cruz, CA 95064, U.S.A.
%\\
}  
\offprints {A. Heger ({\em ahg@mpa-garching.mpg.de})}
\date{Received  ; accepted ,}
\maketitle
%\markboth{A.Heger et al.: \, Pulsations in Red Supergiants}{XXX}

\begin{abstract}
We investigate the pulsational properties of RSG models --- which we
evolve from ZAMS masses in the range $10$ to $20\,\mso$ --- by means
of linear and non-linear calculations. We find period and growth rate
of the dominant fundamental mode to increase with increasing
luminosity-to-mass ratio $L/M$.  Our models obtain relatively large
$L/M$ values due to the inclusion of rotation in the evolutionary
calculations; however, the largest values are obtained at and beyond
central He-exhaustion due to major internal rearrangements of the
nuclear burning regions.  Our non-linear calculations as well as the
behavior of the linear period and growth rate of the pulsations for
periods approaching the Kelvin-Helmholtz time scale of the H-rich
stellar envelope point towards the possibility of large amplitude
pulsations. Such properties are similar to that found in AGB stars and
suggest the possibility of a ``superwind'' to occur before the RSGs
explode as supernovae.  We conclude that changes in global stellar
properties during the last few $10^4\,\yr$ before core collapse may
lead to drastic changes in the pulsational and wind properties of
pre-supernova stars, with marked consequences for the immediate
pre-supernova structure of the star and the circumstellar medium. We
compare our results with observations of long-period OH/IR variables
%in the Galaxy,
% and the LMC, 
and discuss observational evidence for
our scenario from observed supernova light curves, spectra and remnants.

% actually, none of these keyword -- expect for the supernovae -- are in 
the
% thesaurus list...
\end{abstract}

%%%%%%%%%%%%%%%%%%%%%%%%%%%%%%%%%%%%%%%%%%%%%%%%%%%%%%%%%%%%%%%%%%%%%%%

\section{Introduction}
\label{sec:intro}

Massive stars,  which develop a collapsing iron core at the
end of their evolution, are the main agents of nucleosynthesis and
chemical evolution in the Milkyway and other galaxies (cf. Timmes et
al.  1995). According to the standard picture, based on 
stellar evolution calculations
(e.g. Schaller et al. 1992) and on the interpretation of Type II
supernova light curves (Eastman et al. 1994), 
most pre-supernova
stars are red supergiants (RSG) with massive H-rich envelopes.
%this is the result of most stellar evolution calculations
%(e.g. Schaller et al. 1992) and of the interpretation of Type II
%supernova light curves (Eastman et al. 1994). 
In the present paper, we
 focus on a property of the structure of the majority of
pre-supernova stars which is usually neglected: they are pulsationally
unstable. In particular, we suggest the 
possibility for this instability to become violent several
$10^4\,\yr$ before the supernova explosion, with prominent
consequences for the envelope structure of the exploding star as well
as the distribution of the circumstellar matter at the time of the
supernova event.

Observations of long-period variables (LPV) in the Galaxy and
particularly in the Magellanic Clouds (Wood et al. 1983, 1992;
Whitelock et al. 1994) reveal two distinct groups of pulsating cool
stars: intermediate mass stars on the Asymptotic Giant Branch (AGB)
with zero-age main-sequence (ZAMS) masses $\Mz \simle 8\,\mso$, and
RSGs, massive stars with $\Mz \simgr 8\,\mso$.  While the periods of
the visible LPVs are generally shorter than $1000$ days, longer
periods are found in OH/IR sources (Engels et al. 1983; Le Bertre
1993; Jones et al. 1994).
%The similarities in the OH maser
%emission and the IR properties with O-rich Miras (e.g. AGB stars) 
%and RSGs has led to the idea that such sources are composed of a 
%central star which is a long-period variable of late-type, 
%enshrouded by optically thick circumstellar
%dust shells. The formation of such shells is connected to strong mass loss 
%of the central star.  
Most of the OH/IR sources seem to lie on the period-luminosity
relation observed for AGB stars (Whitelock et al.  1991; Feast et
al. 1989; Groenewegen and Whitelock 1996).  Interestingly enough, some
of them show bolometric luminosities much higher than expected for
Miras. One may thus attempt to associate these sources with RSGs.
%though  no extensive 
%theoretical nor observational study allow yet a derivation of a 
%period-luminosity relationship.
However, known pulsating RSGs show generally small amplitudes whereas
the long period OH/IR sources are usually associated to high
amplitudes, as expected for AGB stars (cf. Wood et
al. 1992). Therefore, it is puzzling that the most luminous
OH/IR-sources found by Le~Bertre (1993) show in fact very large
amplitudes.

Recently, a theoretical analysis of RSG evolution and pulsation in the
Large Magellanic Cloud (LMC) has been performed by Li and Gong (1994).
By coupling linear stability analysis and evolutionary models, they
could follow the evolution of stars from $15$ to $30\,\mso$ in the
period-luminosity diagram.  They concluded that RSG models can
reproduce the observed period-luminosity ($P$-$L$) relationship of
LPVs in the LMC, which is, however, on the RSG branch restricted to
periods below $\sim 900$ days.
%up to P $\sim 900$ days. 
%However, they found that the
%theoretical instability strip is cut off at high luminosity, 
%due to the effect of mass loss undergone by their models, 
%leaving the question of overluminous OH/IR sources of 
%periods P $\ge 1000$ days and high amplitudes open.

In the present paper, we reinvestigate the pulsational properties of
RSG models. We compute stellar evolutionary models for massive stars
which evolve into RSGs, and then analyze their linear pulsation
properties. We are using a hydrodynamic stellar evolution code which
allows us to also compute models in the non-linear regime of the
pulsations. Furthermore, we included the physical effects of stellar
rotation on the evolution, which results in a remarkable increase of
the $L/M$ ratio of our models already on the main sequence.  More
dramatically, the $L/M$ ratio increases at and beyond central helium
exhaustion, as it also does for non-rotating stars.  We show that this
property leads to a strong increase of the growth rate of the
fundamental mode, which may have considerable consequences for the
pre-supernova structure of the stars and of the circumstellar
matter. We compare our results with recent observations of luminous
OH/IR sources and with supernova observations.
 
%In the same vein, we present in this letter, the results of 
%linear and non-linear analysis of 
%massive stars with solar metallicity. Evolutionary 
%models, presented in section 2, include the effect of mass loss 
%and rotation which leads to high $L/M$ ratio at the end of the 
%pre-supernova evolution.
%As shown by the non-linear models presented in section 3 and 
%the linear stability analysis (section 4),  the models develop 
%a strong instability as luminosity increases and mass 
%decreases along evolution, with periods above
%1000 days and high growth rates. Comparison of our results with 
%the period - luminosity determined for OH/IR sources is made in 
%section 5. The similarity of the predicted pulsational 
%properties with some OH/IR sources characteristics is most striking. 
%Discussion and conclusion follow in section 5.
 
\section{Stellar evolutionary models}
%%%%%%%%%%%%%%%%%%%%%%%%%%%%%%%%%%%%%%%%%%%%%%%%%%%%%%%%%%%%%%%%%%%%%%%
%	Numerical method and input physics
%%%%%%%%%%%%%%%%%%%%%%%%%%%%%%%%%%%%%%%%%%%%%%%%%%%%%%%%%%%%%%%%%%%%%%%
%\subsection{Numerical method and input physics}
\label{sec:stelev}

The stellar evolution calculations presented here are obtained with an
implicit hydrodynamic stellar evolution code (cf. Langer et al. 1988).
Convection according to the Ledoux criterion and semiconvection are
treated according to Langer et al. (1983), using a mixing length
parameter $\aMLT =1.5$ and semiconvective mixing parameter of
$\aS=0.04$ (Langer 1991). Opacities are taken from Alexander and
Fergusson (1994) for the low temperature regime, and from Iglesias and
Rogers (1996) for higher temperatures.  A parameterization of the mass
loss rate according to Nieuwenhuijzen and de~Jager (1990) is adopted.

During the last decade, the evidence that rotation and rotationally
induced mixing in particular affects the evolution of massive main
sequence stars has grown substantially (cf. Fliegner et al. 1996, and
references therein).  In the present calculations, effects of rotation
on the stellar structure as well as rotationally induced mixing
processes are included in a way similar to Pinsonneault et
al. (1989). The main effect of rotation in relation to the pulsational
properties of the RSGs is an increase of the mass of the
He-core, as well as an enrichment of the envelope with products
of central H-burning.  Both effects are qualitatively discussed in
Langer et al. (1997).  Note that the latter in particular leads to
agreement with the surface abundance pattern observed in OB
main-sequence stars (cf. Fliegner et al. 1996).  The main effect of
the increased core masses is a larger luminosity during the advanced
evolution, an effect which is produced similarly by efficient
convective core overshooting (e.g., Stothers and Chin 1992).
%inclusion of rotation
%is not essential for the pulsational properties of the stellar models 
%discussed in this paper. 
The dynamical effect of rotation on the structure is found to be
negligible and does not affect the pulsational properties of our RSG
models.  The ratio of rotation velocity to the critical value at the
surface is below $\sim 0.3\,\%$ for all RSG models considered.
 
%%%%%%%%%%%%%%%%%%%%%%%%%%%%%%%%%%%%%%%%%%%%%%%%%%%%%%%%%%%%%%%%%%%%%%%
%	The evolutionary sequences
%%%%%%%%%%%%%%%%%%%%%%%%%%%%%%%%%%%%%%%%%%%%%%%%%%%%%%%%%%%%%%%%%%%%%%%
%\subsection{The evolutionary sequences}
%\label{sec:evseq}

We compute four evolutionary sequences for different ZAMS masses,
$10$, $12$, $15$ and $20\,\Msun$, with an initial composition of $Y=0.28$
and $Z=0.02$ (cf. Table~1).  We adopt an initial equatorial rotation
velocity of about $200\,\kms$, which is a typical value for these
stars (Fukuda, 1982; Howarth et al., 1997) 
and corresponds to $\sim 35\,\%$ of their
critical rotation speed.  We start the stellar evolution with a fully
convective, rigidly rotating model on the pre-main sequence, and we
follow it beyond central neon exhaustion so that the stellar
envelope can safely be considered to be in the pre-supernova state.
The evolutionary tracks of our models in the Hertzsprung-Russell
diagram (HRD) are shown in Fig.~1, starting at 
H-ignition.

%%%%%%%%%%%%%%%%%%%%%%%%%%%%%%%%%%%%%%%%%%%%%%%%%%%%%%%%%%%%%%%%%%%%%%%
\begin{figure}
  \epsfxsize=88mm
  \epsfysize=88mm
%  \epsfbox{/afs/mpa-garching.mpg.de/home/ahg/Puls/hrd.eps}
    %%\epsfbox{fig1new.ps}
\epsfbox{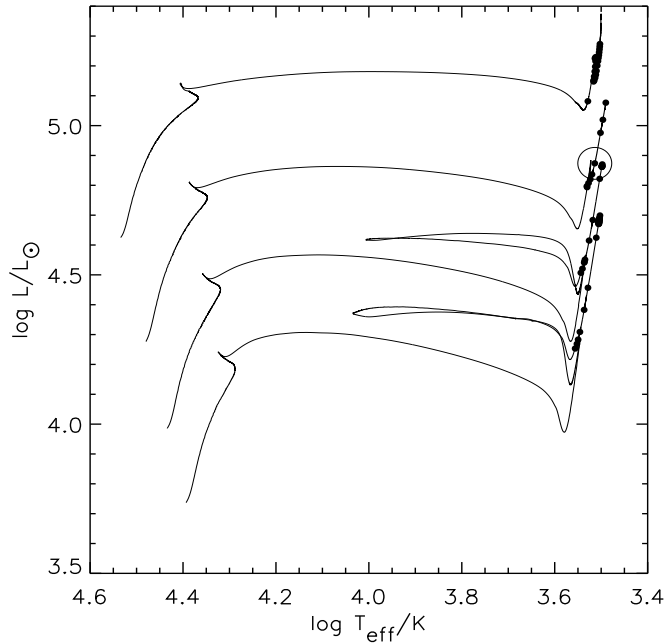}
  \caption{Evolutionary tracks of the $10$, $12$, $15$ and $20\,\Msun$
           sequences from the ZAMS to central neon exhaustion.  Models
           for which a linear pulsation analysis has been performed
           (cf. Sect.~3) are marked by dots.  The circle marks the
           initial model investigated in Sect.~4; cf. Figs.~3 and 5.}
  \label{fig:hrd}
\end{figure}
%%%%%%%%%%%%%%%%%%%%%%%%%%%%%%%%%%%%%%%%%%%%%%%%%%%%%%%%%%%%%%%%%%%%%%%

%%%%%%%%%%%%%%%%%%%%%%%%%%%%%%%%%%%%%%%%%%%%%%%%%%%%%%%%%%%%%%%%%%%%%%%
\begin{table}
  \caption{Time-scales of hydrogen burning, helium burning, and
   life time at $\Teff < 5000\,$K. Furthermore, 
  properties of the last computed model of each sequence
  are given (cf. Fig.~1):
  surface helium mass fraction, luminosity, effective
  temperature, radius, final stellar mass, He- and
  C/O-core mass
%, time on spent as RSG before supernova explosion,
%models taken for the stability analysis 
%(first and last full circles on the tracks in Fig.~1),
 and Kelvin-Helmholtz time scale of the H-rich envelope}
  \label{tab:evpar}
  \begin{center}
  \begin{tabular}{lrrrr}
    \hline
    \noalign{\smallskip}
    ZAMS mass / $\Msun$ & $10$ & $12$ & $15$ & $20$ \\
    \noalign{\smallskip}
    \hline
    \noalign{\smallskip}
%      age / 10$^6\,\yr$ &     &     &     &        \\
%    $\vvc$ & $0.0032$ & $0.0013$ & $0.0011$ & $0.0003$ \\
     $\thyd/10^6\,$yr & 25.5 & 18.9 & 13.8 & 9.6 \\
     $\thel/10^6\,$yr & 2.86 & 1.99 & 1.22 & 0.64 \\
     $\tred/10^6\,$yr & 0.39 & 0.46 & 1.23 & 0.66 \\
    \noalign{\smallskip}
    \hline
    \noalign{\smallskip}
     $\Hef$ & $0.39$ & $0.40$ & $0.42$ & $0.45$ \\
     $\log(L/\Lsun)$ & $4.66$ & $4.86$ & $5.05$ & $5.36$ \\
     $\te/\K$ & $3203$ & $3152$ & $3107$ & $3164$ \\
     $R/\rso$ & $702$ & $903$ & $1161$ & $1614$ \\
     $M/\Msun$ & $9.2$ & $10.4$ & $10.9$ & $11.0$ \\
%     $M_{env}/\Msun$ & $6.4$ & $6.7$ & $5.8$ & $3.3$ \\ 
     $\MHe/\Msun$ & $2.8$ & $3.6$ & $ 5.1$ & $7.7$ \\
     $\MCO/\Msun$ & $1.8$ & $2.3$ & $3.4$ & $4.6$ \\
%     $\tRSG/10^5\,\yr$ & $3.9$ & $3.5$ & $6.9$ & $6.4$ \\
     $\tKH/\yr$ & $28.7$ & $17.1$ & $7.7$ & $1.5$ \\ 
    \noalign{\smallskip}
    \hline
  \end{tabular}
  \end{center}
\end{table}
%%%%%%%%%%%%%%%%%%%%%%%%%%%%%%%%%%%%%%%%%%%%%%%%%%%%%%%%%%%%%%%%%%%%%%%

The tracks presented in Fig.~1 are qualitatively similar to those
obtained by previous calculations including overshooting (Schaller et
al. 1992, Stothers and Chin 1992).  The total mass lost by our stars
is dominated by RSG winds. Due to a slightly larger luminosity during
central He-burning compared to the tracks of Schaller et al. (1992),
we obtain somewhat smaller final masses (cf. Table~1); they are
comparable to those obtained by Meynet et al. (1994).  The effect of
rotation on the final $L/M$ ratio can be summarized as follows: it
results in an increase of the luminosity due to a larger He-core, and
the correspondingly higher mass loss rate yields a smaller final mass;
both properties translate into a higher $L/M$ value, which is a key
parameter for the pulsations (cf. Sect.~3).
  
%The first ``dredge-up'' after the star has left the MS
%enriches the surface layers to between $39\,\%$ and $45\,\%$ of
%helium.  The second dredge-up at the end of central
%He-burning leaves these values almost unchanged.  The two
%lower mass models ($10$ and $12\,\Msun$) undergo a ``blue-loop'' at
%about $45\,\%$ and $33\,\%$ central He, respectively.

%We stress again that $L/M$-ratios in the RSG stage
%can as well be obtained through convective core overshooting
%instead of rotation, 
%depending on the assumed amount of overshooting (REFERTENCES).
%It is beyond the scoop of the present letter to enter into a debate on the 
%effects of rotation versus core overshooting and on the confrontation of 
%both 
%effects with observations. Detailed discussion on the problem can be found 
%in 
%In the present work, rotation has to be regarded as a  possible
%mechanism for additional mixing
%processes, supported by the fact that stars do rotate.  As already 
%mentioned,
%the dynamical effects of rotation {\it in the RSG phase}  are 
%rather negligible, due to momentum angular loss and radius extension. 

%%%%%%%%%%%%%%%%%%%%%%%%%%%%%%%%%%%%%%%%%%%%%%%%%%%%%%%%%%%%%%%%%%%%%%%
%	LINEAR stability analysis
%%%%%%%%%%%%%%%%%%%%%%%%%%%%%%%%%%%%%%%%%%%%%%%%%%%%%%%%%%%%%%%%%%%%%%%
\section{Linear stability analysis}
\label{sec:stalin}

We perform a linear non-adiabatic stability analysis of the RSG
models marked by dots in Fig.~1.
%For the two lower-mass stars ($10$
%and $12\,\Msun$) only models after the blue loop in the HRD
%are considered.
For the two lower-mass sequences ($10$ and $12\,\Msun$), the stability
analysis is started after they returned from their blue loop 
(cf. Fig.~1)
near the end of central He-burning. 
%($\sim 10\,\%$ helium left in the core). 
For the $15$ ($20$)
$\msol$ models, the analysis begins at a
central helium mass fraction of $0.43$ ($0.97$).  
%
%The time spent by the stars between the first and last model analyzed
%is indicated in Table~1 ($\tRSG$) for each ZAMS mass, and is about
%$5$ $10^5\,\yr$.
%
The radial pulsation code used for this study is
described in Jeannin et al. (1996). Our analysis is based on 
complete stellar models i.e.  the whole stellar structure from the
center to the surface is taken into account.
The convective flux is assumed to be frozen in for 
the stability analysis 
%though convective energy transport in the envelope is efficient
(cf. Sect.~4).
 
\begin{figure}
  \epsfxsize=88mm
  \epsfbox{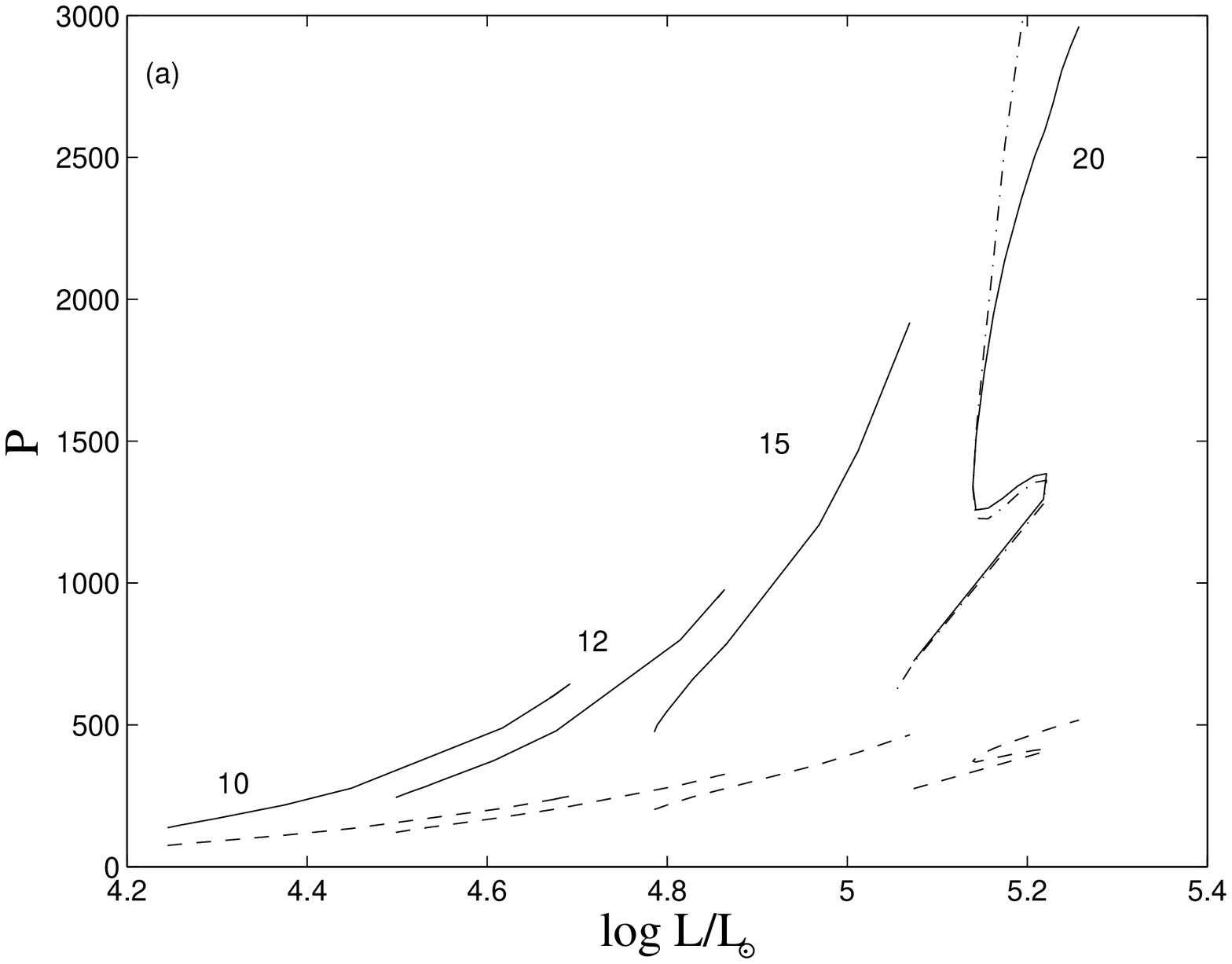}
  \epsfxsize=88mm
  \epsfbox{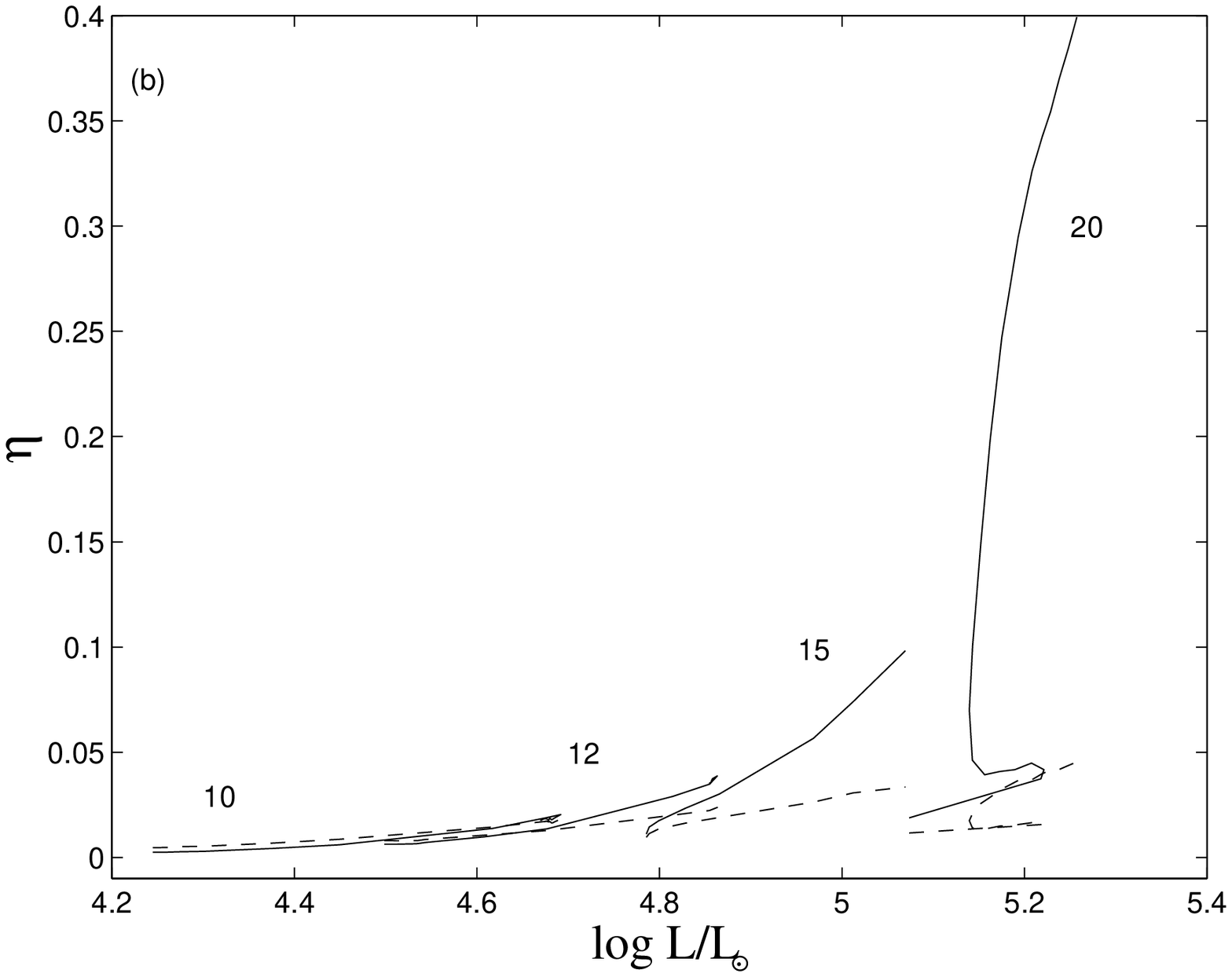}
  \caption[ ]{Period in days ({\bf a}) and growth rate ({\bf b}) of
              the pulsations found in the stellar models of the $10$,
              $12$, $15$ and $20\,\mso$ sequences marked in
              Fig.~1. The full drawn lines correspond to the
              fundamental mode, the dashed lines to the first
              overtone.  The dash-dotted line in (a) corresponds to
              the adiabatic period of the $20\,\mso$ sequence.}
  \label{fig:P,eta-L}
\end{figure}

\begin{figure}
\epsfxsize=88mm
\epsfysize=60mm
\epsfbox{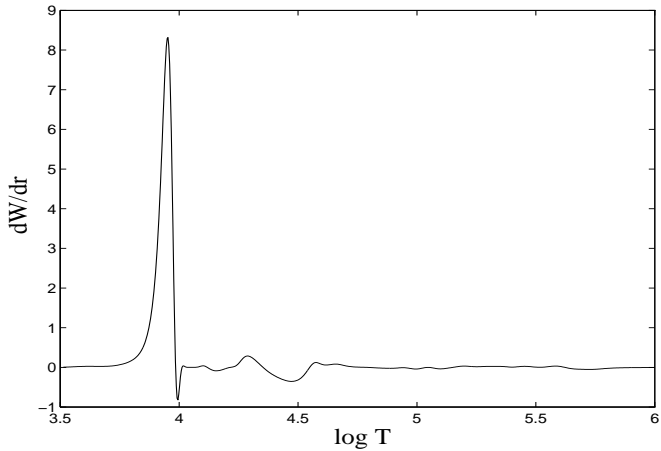}
\caption[ ]{ Differential work (in arbitrary units) as a function of temperature for a
model of the 15 $\msol$ sequence characterized by $\log L/\lsol = 4.89$, $\te =
3260\,\K$ and $M \simeq 11\,\msol$, and indicated by a circle in Fig. 1. 
Excitation zones correspond to dW/dr $>$ 0.}
  \label{fig:work}
\end{figure}

The linear pulsation code assumes that the stellar models are in
complete thermal and hydrostatic equilibrium. Although the
evolutionary models include rotation and slightly depart from thermal
equilibrium, we have checked the consistency of this approximation.
%We verify that the dynamical time scale of the models
%remains much shorter than the thermal and rotational
%time scales.
Rotation has no dynamical effect on the RSG pulsations.
As the stellar envelope expands when the star becomes a RSG
it spins down dramatically. Furthermore, as the oscillations take place 
 only in the envelope, they are not affected by the
rotation of the helium core.  
For each
model at a given mass, $L$ and $\te$, we construct static envelope
models with the same input physics (opacity, mixing length,
composition).  The pulsation properties of the envelopes are then
compared to those of the corresponding evolutionary model. In all
cases, the differences are less than $5\,\%$ for the period.  The
analysis was stopped for the brightest models, when the decreasing
evolutionary time scale starts to invalidate our approximation.

%\begin{figure}
% \epsfxsize=88mm
% \epsfysize=98mm
% \epsfbox{fig7b.ps}
%
% \caption[ ]{Growth rate of the fundamental mode as a function of the
%    the $L/M$ ratio, for the RSG models of the four computed evolutionary
%    sequences (cf. Figs.~1 and~2).
%   Small circles designate the time when the central helium mass fraction
%   has reached 1\%, i.e. roughly central helium exhaustion.}
%
%  \label{fig:eta-L/M}
%
%\end{figure}

%As already mentioned, the convective flux is frozen in in the stability 
%analysis, though
%convective energy transport in the envelope is efficient. This is 
%certainly a
%poor approximation to the models, due to the lack of a coherent theory
%which describes the coupling between convection and acoustic
%waves (cf. \S 3).

The periods P
and relative growth rates $\eta$
(defined as minus the ratio of the imaginary part of the
eigenfrequency $\sigma$ to its real part, adopting a time dependence of
the form exp(i$\sigma$t)) 
of the two lowest-order modes of the four stellar sequences are displayed
in Fig.~2.  Typical values of the period for the fundamental
mode range between $200$ and $3000$ days.  The periods found for the
first overtone do not exceed $500$ days in any case.  Both modes are
unstable ($\eta > 0$) for all investigated models, and the growth rates
%(defined as minus the ratio of the imaginary part of the
%eigenfrequency to the real part) 
increase with luminosity for a given ZAMS mass.  The
higher the ZAMS mass, the higher the maximum value of the growth rate
reached.
%{\bf or $L/M$ values ... to be changed  N.}
The growth rate derived for the fundamental mode is always larger than
that of the first overtone.  The latter remains systematically smaller
than $0.05$, whereas a maximum value of $0.4$ for the fundamental mode
is obtained in the $20\,\msol$ case.  For
the $15\,\mso$ model, the fundamental mode growth rate increases from
$\sim 0.01$ to $\sim 0.1$.  It remains smaller than $0.04$ for
the $10$ and $12\,\mso$ stars (cf. Fig.~2).

%it reaches a value of $0.4$ in the most
%extreme case, indicating that the growth time of the pulsation is
%of the same order as the pulsation period.
The excitation mechanism appears to be related to the traditional
${\kappa}$ mechanism as indicated by the work integral W of the
models. The driving of the pulsation is essentially provided in the
hydrogen ionization zone. This is illustrated in Fig. 3 where dW/dr is
displayed as a function of temperature for a model of the 15 $\msol$ sequence
at the end of central He-burning. 
%We did not find any evidence for the occurrence of
%strange modes characterized by complex conjugate eigenvalues
%(cf. Gautschy \& Glatzel 1990) and which seem to appear in hotter
%models (see Glatzel and Kiriakidis 1993; Aikawa 1993).

\begin{figure}
  \epsfxsize=88mm
  \epsfysize=60mm
  \epsfbox{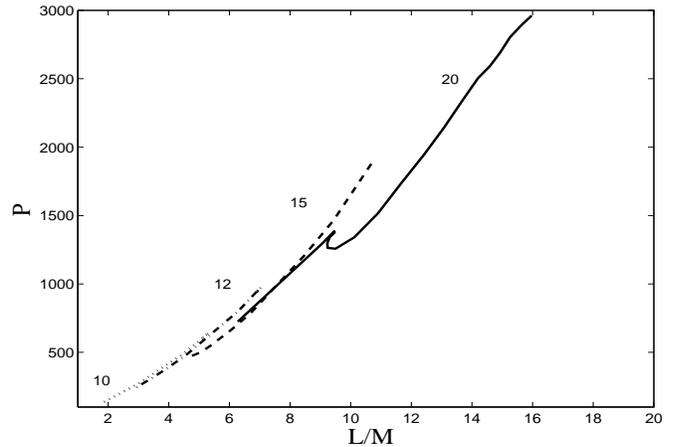}
  \caption[ ]{Period of the fundamental mode (in days) as a function
              of the ratio $(L/M)$ (in units of $10^3\,\Lsun/\Msun$).
              The different curves correspond to different ZAMS masses
              as indicated.}
  \label{fig:L/M-P}
\end{figure}

The large value of the periods and growth rates mentioned above can be
related to the large $L/M$ values obtained from the evolutionary
calculations.  This behavior is illustrated in Fig.~4
%and \Figx{eta-L/M} 
which shows the variation of $P$ 
%and $\eta$, respectively,
with $L/M$ for the four investigated sequences. The variation $P
\propto L/M$ is connected to the properties of convective pulsating
{\it envelopes} which obey a period-radius relation $P\propto R^2/M$
rather than the canonical $P \propto R^{3/2}/M^{1/2}$. This has been
shown analytically by Gough et al. (1965) for polytropic envelopes of
index $n=3/2$ corresponding to a fully convective structure. Since the
effective temperature varies hardly on the RSG branch, the variation
$P\propto R^2/M$ is equivalent to $P \propto L/M$. 

The relevant
quantity for the growth or decay of pulsations, connected to the
growth rate, is the Kelvin-Helmholtz or thermal time scale of the
H-rich envelope $\tKH \propto M M_{\rm env} / L R$.  It governs the
degree of non-adiabaticity in the envelope and thus the heat exchange
with the pulsation.  In our case, $\tKH$ decreases as the models get
brighter (and lose mass), and for the brightest models
$\tKH$ becomes very close to the fundamental
period (cf. Fig.~7 below).  
Non-adiabatic effects are thus important and are responsible
for the high values of the growth rates found as the luminosity
increases and $\tKH$ decreases (see also Ostlie and Cox, 1986).

The large fundamental to first overtone period ratio
$P_0/P_1$ displayed by the models in Fig.~2 can be directly
related to the internal structure of the stars. With increasing
luminosity, the star tends toward a dynamically unstable configuration
with the mean value of the adiabatic exponent $\Gamma\!_1 \sim 4/3$. In
this case, the solution of the linear adiabatic wave equation (cf. Cox
1980) shows that the square of the adiabatic pulsation frequency
becomes negative and the adiabatic period becomes mathematically
infinite.  The divergence of the adiabatic period for the fundamental
mode is illustrated in Fig.~2a for the $20\,\mso$ sequence.  As long as the
non-adiabatic effects remain small, the non-adiabatic period is rather
close to the adiabatic value and follows the same behavior.  When the
non-adiabatic effects become strong however, the non-adiabatic period
start to depart from the adiabatic solution (cf. Fig.~2a).

\section{Non-linear computation of the pulsations}
\label{sec:stanon}

As mentioned above, the stellar evolution models are computed with a
hydrodynamic code, which is therefore able to follow instabilities on
dynamical time scale. Usually, those are not found due to the large
evolutionary time step compared to the dynamical time scale of the
star. However, during the late nuclear burning phases (C- and
Ne-burning), nucleosynthesis imposes an extremely small time step
comparable to or even smaller than the dynamical time scale. I.e.,
only the combination of hydrodynamic stellar evolution calculations of
mass losing RSGs in late nuclear burning stages enabled us to find the
strong instability described here,

Fig.~5 displays, in the HRD, the growth of the pulsational
instability of a model of the $15\,\Msun$ sequence at the end of
central He-burning ($0.2\,\%$ helium left in the center).  The initial
model for the non-linear pulsation analysis, indicated by a circle in
Fig.~1, is characterized by $\log L/\lsol = 4.89$, $\te =
3260\,\K$ and $M \simeq 11\,\msol$.  The first $50$ pulsation cycles
give a period P $\simeq 800$ days and an estimated growth rate $\eta =
0.033$, corresponding to the fractional increase of the radius
amplitude per pulsation cycle.
%by a factor of
%$\exp{2\pi\eta}$.  
These values are in excellent agreement with those found for the
fundamental mode by the linear stability analysis performed on the
equilibrium model ($P_0 = 802$ days and $\eta_0 = 0.033$). 
Fig.~5 shows that after 75 cycles, 
the luminosity amplitude has already grown to a
factor of two, i.e. to $0.75\,\Mag$ in $\Mbol$.
%$0\magp 75$ magnitudes in $\Mbol$.

%%%%%%%%%%%%%%%%%%%%%%%%%%%%%%%%%%%%%%%%%%%%%%%%%%%%%%%%%%%%%%%%%%%%%%%
\begin{figure}
  \epsfxsize=88mm
  \epsfysize=88mm
  \epsfbox{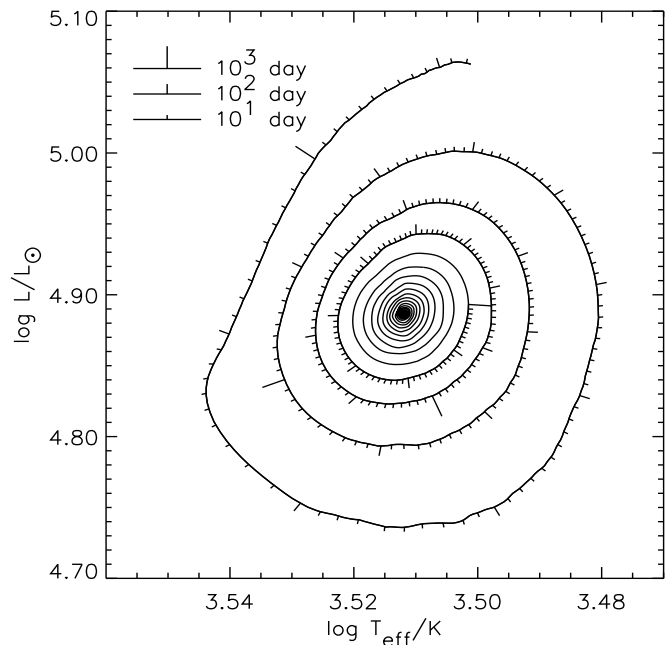}
  \caption{Growing pulsational instability of a $\sim 11\,\Msun$ RSG
  model (evolved from a $15\,\Msun$ ZAMS star) at the end of central
  He-burning, traced by our stellar evolution code.  The period of the
  oscillation is about $800$ days, the Kelvin-Helmholtz time scale of
  the stellar envelope is roughly $30\,\yr$. About 75 pulsation cycles
  have been followed.}
  \label{fig:spiral}
\end{figure}
%%%%%%%%%%%%%%%%%%%%%%%%%%%%%%%%%%%%%%%%%%%%%%%%%%%%%%%%%%%%%%%%%%%%%%%%

Our non-linear calculations are limited for two reasons.
The first is that our code is fully implicit and therefore subject to
strong numerical damping (cf. Appenzeller 1970). Thus, we are not able
to detect the pulsations predicted by the linear theory
(Sect.~3) for the case of small growth rates (e.g. for the
$10\,\mso$ sequence).  
%On the other hand, we are limited 
%when going to  relatively large
%amplitudes which can generate shocks and for which our code is not 
%adapted.
Secondly, we are limited on the other extreme to not too large
amplitudes,
%such that the pulsation velocities remain below the local
%sound speed, 
since our code is not set up to deal with shock waves.  For this
reason we did stop the calculation shown in Fig.~5 when the surface
velocity reached about $10\,\kms$, and we did not attempt to compute
non-linear pulsations for increasing amplitudes.

The main uncertainty in the description of the pulsational
behavior of our models is certainly due to the unknown feedback of
convection.  As in Miras, the convective and the pulsation time scale
are comparable, and at present no solution to the coupled problem
exists (cf. Gautschy \& Saio 1995).  In contrast to the linear
stability analysis presented in Sect.~3, where the convective
flux is frozen in, it is assumed to adjust instantaneously in the
hydrodynamic stellar evolution code.  Both treatments thus represent
simple but different possibilities to deal with the convective flux
feedback and  give essentially the same
periods and growth rates.  Even within a ``phase-lag'' approach
(Arnett 1969) to describe time-dependent convection, Langer (1971) has
shown that the general pulsation properties (period, growth rate)
derived from linear stability analysis do not depend strongly on the
phase-lag parameters. However,
the growth rates derived in the present work, as they depend on
the energy transport, should only be taken as indicative.
A physically correct 
treatment of thermal and dynamical coupling of convection and
pulsation --- which is not yet available --- 
may be necessary to predict them reliably.
We can be more confident on the values of the period, which
essentially depend on the structure and the sound speed in the
envelope.

%This result may not allow to advance general
%conclusions concerning the effect of convection on pulsation,
%It could as well suggest the
%limitation of such a rough treatment to 
%take into account correctly the thermal
%and dynamical coupling between convection
%and pulsation (Gonczi and Osaki, 1980).
%and the growth rates derived in the present work, as they depend on
%the energy transport, should only be taken as indicative.
%We will therefore mainly concentrate on their
%general behavior with  stellar parameters (see next section).
%We may remain more confident on the values of the period, which
%essentially depend on the structure and the sound speed in the
%envelope.  In any case, we want to emphasize that more work towards a
%physically correct thermal and dynamical coupling of convection and
%pulsation is certainly necessary (cf. Gonczi and Osaki, 1980).

The results of the present analysis and the fate of the star 
thus remain uncertain, since we ignore the effects of convection on the
pulsational driving. Nevertheless, the example shown in Fig.~5 and the similarity of
the pulsational properties (linear and non-linear) discussed above with those obtained for AGB
stars, in particular the huge increase in period and growth rate
during the final evolutionary stages, may suggest
the possibility of pulsations of large amplitude in RSGs in connection
with a strong increase in the mass loss rate, i.e. a ``superwind''
phase as that observed for Miras and expected for the terminal AGB evolution. This will be
discussed in the next Section.

%Even more so, the diverging character of the period and growth
%rate with increasing luminosity shown in \Figx{P,eta-L} which can be
%understood as the approach to a dynamical runaway for situations where
%the period approaches the Kelvin-Helmholtz time scale of the stellar
%envelope (cf. Fig.~6), 
%
%The similarity of the pulsational properties discussed above with 
%those obtained for AGB stars suggests the possibility for the RSG models
%to undergo strong mass loss, like the ``superwind'' event.
%This will be examined in the next Section.

\section{Discussion}
\label{sec:discus}

Fig.~6 shows the $P$-$L$ relation derived from our models in
comparison to observations of long period OH/IR sources from Engels et
al. (1983), Le Bertre (1991, 1993) and Jones et al. (1994).  The mean
$P$-$L$ relation observed for O-rich Miras (Feast et al. 1989)
is also shown.  Though most of the OH/IR sources follow the latter
$P$-$L$ relationship and may thus be identified as AGB stars, some
objects, mostly from Le Bertre (1993), are overluminous and fall on
the curves of our RSG models. Note that  the narrow relation
between the period and the $L/M$-ratio found in Fig.~4 might provide an 
interesting tool to deduce actual masses of these sources.

\begin{figure}
 \epsfxsize=88mm
 \epsfysize=80mm
 \epsfbox{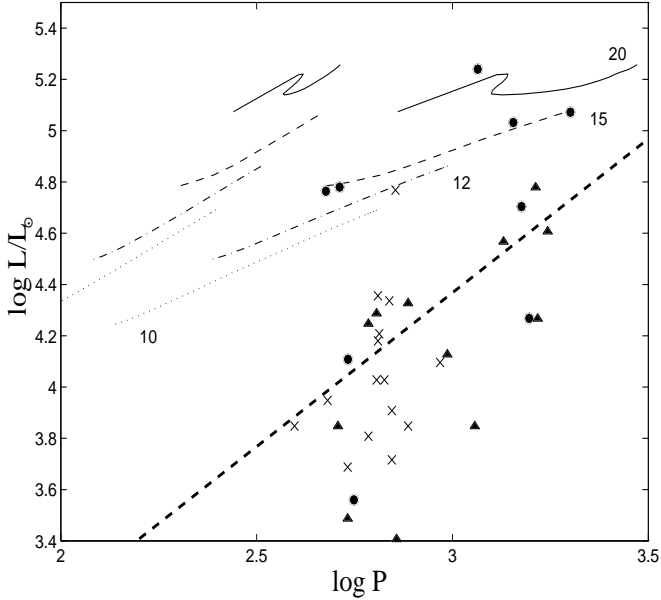}
 \caption[ ]{ Evolution of the luminosity as a function of the period
    (in days) for RSG models along the different calculated tracks,
    for the fundamental mode (right curves) and the first overtone
    (left curves). The ZAMS masses are indicated for the fundamental
    periods: $10\,\msol$ (dot), $12\,\msol$ (dash-dot), $15\,\msol$
    (dash) and $20\,\msol$ (solid line). The thick dashed line
    corresponds to the relationship observed for O-rich Miras
    (Feast et al. 1989).  Observations of OH/IR sources are from:
    Engels et al. (1983, triangles), Le Bertre (1991, 1993, full
    circles) and Jones et al. (1994, crosses).}
  \label{fig:L-P}
\end{figure}
 
Although the overluminous sources show large amplitudes ($1$ to
$3\,\Mag$), a feature which is generally attributed to AGB stars, the
RSG nature of at least some of them appears to be consistent with our
results, which suggests that RSGs may also be able to develop large
amplitude pulsations.  
Most of the overluminous OH/IR sources show extremely strong mass loss
from the central object, 
with rates as high as $10^{-4}\msoy $ (Le Bertre 1991).
This is  in support of 
the idea of high mass loss in connection with large periods,
growth rates and amplitudes. The order of magnitude of the
observed mass loss rates
is similar to what is discussed for the final AGB ``superwind'' 
(Vassiliadis and Wood 1993); 
however, note that Le Bertre's sources are oxygen-rich
while the classical AGB superwind is expected in carbon-rich sources. 

Fig.~7 shows that very large pulsation periods and
consequently large amplitudes and mass loss rates may be expected to
occur at and beyond central helium exhaustion for our model sequences,
over a time scale of some 10$^4\,$yr, implying the loss of most of the
hydrogen-rich envelope which remained on the star so far
(cf. Table~1).  In this context we emphasize the self-enhancing
character of the pulsational mass loss increase since, as the envelope
mass decreases, $L/M$ is increased and the Kelvin-Helmholtz time scale
of the envelope is decreased.  This effect is not yet taken into
account in the present calculations, where only the mass loss
according to Nieuwenhuijzen and de Jager (1990) is
adopted. Consequently, the superwind instability may even develop 
stronger than Fig.~7 may suggest.

\begin{figure}
 \epsfxsize=88mm
 \epsfysize=80mm
 \epsfbox{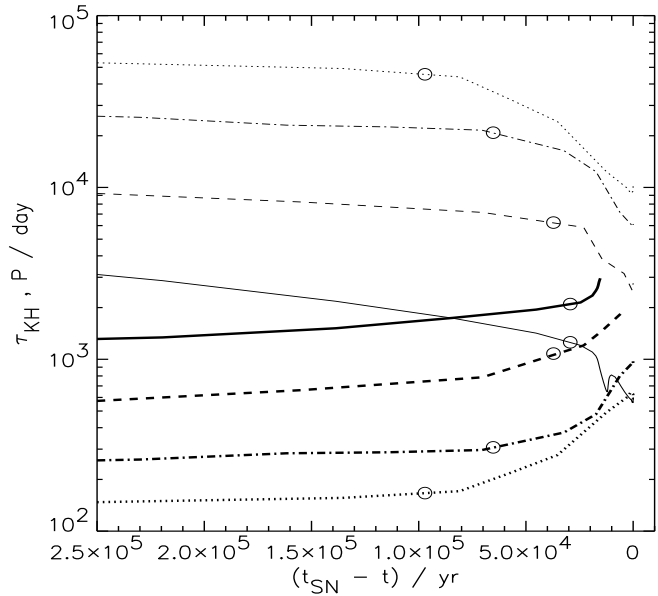}
 \caption[ ]{Kelvin-Helmholtz time scale of the H-rich stellar
   envelope (thin lines) as function of the time left until the
   supernova explosion of the star, for the computed $10$ (dot), $12$
   (dash-dot), $15$ (dash), and $20\,\Msun$ (solid line) sequences
   (cf. Fig.~1).  The thick lines show the period of the
   fundamental mode as derived from the linear stability analysis
   (cf. Sect.~3).  Small
   circles designate the time when the central helium mass fraction
   has reached 1\%, i.e. roughly central helium exhaustion.
   It is evident that, at least for the $15$ and
   $20\,\mso$ sequences pulsation periods of the order of the
   Kelvin-Helmholtz time scale occur. Note that for the last models
   of these two sequences pulsation periods could not be derived due
   to strong departures from thermal and hydrostatic equilibrium. }
  \label{fig:tKH,P-t}
\end{figure}

%Conclusions on the nature of these objects would be premature at present,
%since the determination of their luminosity and periods are rather
%uncertain.
%The main point to stress here is 
%the clear distinction predicted between AGB stars
%and late-type supergiants in a $P$-$L$ diagram, as illustrated 
%in \Figx{L-P}
%Regarding the properties of OH/IR sources, high mass loss
%from the central object is observed, with
%rates as high as $10^{-4}\, \msol$ (Le Bertre 1991), and

The probability of observing RSGs in a stage of large amplitude
pulsation or in a ``superwind'' phase is not very large (cf. Fig.~7);
however, such events might have marked consequences on the appearance
of the supernova explosion.  Weiler et al. (1992) have found a
periodic modulation of the pre-supernova mass loss from the radio
light curve of SN~1979C.  Although in this case the inferred period
in the pre-supernova mass loss
($\sim 4000\,\yr$) does not fit to dynamical pulsations discussed
here, this work shows the potential of radio observations of
supernovae to find evidence for those.  A period of the radio signal
of the order of few days would be expected in the early phase of the
supernova evolution.

A ``superwind'' occurring some $10^4\,\yr$ before the explosion could
reduce the mass of the H-rich envelope of the star to less than
$1\,\mso$; this would make the star He-rich and hotter than a
typical RSG (cf. H\"oflich et al. 1993). 
These features are reminiscent of SN~1993J, which
apparently was a He-enriched K~supergiant with an envelope mass
of less than $\sim 0.5\,\mso$ (Nomoto et al. 1993, Woosley et
al. 1994).  The progenitor structure of this supernova is currently
interpreted as a result of close binary interaction (e.g. Nomoto et
al. 1993); however, no companion star could be yet found. This
situation is similar to that of SN~1987A: there also, the
circumstellar matter clearly indicates that the progenitor star
experienced a major structural change several $10^4\,\yr$ before the
supernova explosion.

Another example of a pre-supernova star which experienced the loss of
most of the H-rich envelope before the explosion appears to be
the progenitor of SN~1054, i.e. the progenitor of the Crab
nebula. H-rich gas associated with this supernova remnant has
only been found recently by Murdin (1994) in the form of an
H-rich halo of several $\mso$ (cf., however, Fesen et al. 1997).

Tsiopa (1995) has summarized evidence from supernova spectra for the
ejection of material some $10^4\,\yr$ before the explosion; she found
evidence in about half a dozen cases, including the well known
SN~1983K and SN~1988Z.

The scenario outlined above has also consequences for the formation of
Wolf-Rayet stars. If the critical $L/M$ value for the occurrence of
strong pulsations and associated strong mass loss would occur in a RSG
already during central helium burning, it would evolve into a Wolf-Rayet
star. This may happen for larger initial masses as those considered by
us, or for larger initial rotation rates. The transition objects may
be the so called yellow hypergiants, highly variable and very luminous
stars evolving off the RSG branch (Smoli\'nski et
al. 1989). Correspondingly, the minimum ZAMS mass for Wolf-Rayet star
formation might be reduced below values currently found in stellar
evolution calculations (i.e. $25 \ldots 35\,\mso$; cf. Schaller et al.
1992, Meynet et al. 1994), thereby increasing the number of single
stars contributing to supernovae of Type~Ib/c (cf. Woosley et
al. 1993, Langer and Woosley, 1996).

In this context it is noteworthy that Garc\'{\i}a-Segura et
al. (1996), on the basis of hydrodynamic computations for the
pre-supernova evolution of the circumstellar material around massive
stars, concluded from the location of the 
quasi-stationary flocculi of the Cas~A
supernova remnant that the progenitor star evolved from
the RSG branch into a Wolf-Rayet star roughly $10^4\,\yr$ before the
supernova explosion. An increased mass loss due to pulsations during
the last few $10^4\,\yr$ in the evolution of RSGs makes
such a scenario even more likely.

\section{Conclusions}
\label{sec:concl}

In the present work, we analyze the pulsational properties of RSG
models. 
%Despite the uncertainty of our results due to the unknown
%effect of the coupling between convection and pulsations, it is
%remarkable that a) our results agree well with the recent work of Li
%and Gong (1994), and b) our linear and non-linear results agree
%concerning the period and growth rate of the pulsations
%despite the quite different treatment of convection in both cases.
Due to the inclusion of rotation in the stellar evolution
calculations, our RSG models evolve to high values of $L/M$.
Consequently, their pulsations are characterized by long periods (up
to 3000~days) and large growth rates, much greater than
previously found in the work of Li and Gong (1994).
We thus argue that RSGs may
display large amplitude pulsations like AGB stars, 
preferentially as
OH/IR sources, and perhaps even evolve a ``superwind''.  Due to the
marked increase in the luminosity of RSGs during the last couple of
$10^4\,\yr$, such a behavior is most likely to happen during the final
evolutionary phases preceding the supernova explosion. Observed properties
of  Type~II supernovae and their remnants, and of overluminous OH/IR
 sources, give some support to this scenario.
%Therefore we
%argue that, although this scenario is somewhat speculative, its
%consequences may have been already observed in Type~II supernovae and
%their remnants.

More detailed and extended analysis, taking into account non-linear
effects occurring at high amplitudes and the possible effects of shock 
waves
% and the coupling
%between pulsations and convection 
should be performed in the future,
in order to determine more precisely
the amplitudes of the pulsation 
and to accurately estimate the region in which RSG pulsations are
important for the evolution of the stars and the pre-supernova
circumstellar matter.

\begin{acknowledgements}
We are grateful to Jens Fliegner for implementing the rotation physics
into the stellar evolution code and to Achim Weiss for providing us
with updated opacity tables.  We thank Thibaut Le Bertre and Stan
Woosley for useful discussions.  This work has been supported in part
by a ``DAAD-\-Dok\-to\-ran\-den\-sti\-pen\-di\-um aus den Mit\-teln
des 2. Hoch\-schul\-pro\-gramms'', by NSF grant AST~94-17161 and by
the Deutsche Forschungsgemeinschaft through grant La~587/8-2.
\end{acknowledgements}

\end{document}